%% file: main.tex
\documentclass[conference]{IEEEtran}
\IEEEoverridecommandlockouts
\usepackage{cite}
\usepackage{amsmath,amssymb,amsfonts}
\usepackage{algorithmic}
\usepackage{graphicx}
\usepackage{textcomp}
\usepackage{xcolor}
\def\BibTeX{{\rm B\kern-.05em{\sc i\kern-.025em b}\kern-.08em
    T\kern-.1667em\lower.7ex\hbox{E}\kern-.125emX}}
    
\usepackage{comment}
\usepackage{multirow}
\usepackage{xcolor}
\usepackage{amsmath}
\usepackage{xspace}
\usepackage{tikz}
\usepackage{pgfplots}
\usetikzlibrary{patterns}
\usetikzlibrary{shapes,arrows}
\usetikzlibrary{calc, intersections}
\usetikzlibrary{mindmap,trees}

\begin{document}

\title{Correlating Effectiveness of Pointer Analysis Techniques with Patterns in Embedded System Code}

\author{\IEEEauthorblockN{Komal Pathade}
\IEEEauthorblockA{\textit{TCS Research} \\
Pune, India \\
komal.pathade@tcs.com}
}

\maketitle

\begin{abstract}
A pointer analysis maps the pointers in a program to the memory locations they point to. 
In this work, we study the effectiveness of the three flavors of pointer analysis namely \emph{flow sensitive, flow insensitive,} and \emph{context sensitive} analysis on seven embedded code sets used in the industry. We compare precision gain i.e., the reduction in the number of spurious memory locations pointed by a pointer in each of these settings. 
We found that in 90\% of cases the pointer information was same in all three settings. In other cases, context sensitive analysis was 2.6\% more precise than  flow sensitive analysis which was 6.8\% more precise than flow insensitive analysis on average. We correlate precision gain with coding patterns in the embedded systems\textemdash which we believe to be first of its kind activity.

\end{abstract}


\section{Introduction}\label{sec:Introduction}
Embedded systems are an important part of today's life, specific examples include\textemdash controlling devices in car, avionics in aircraft, telematic systems for traffic control, pacemakers, and control systems in nuclear reactors. In the embedded systems where safety and reliability is critical, early defect detection can prevent serious destruction. Hence, for safety critical and reactive systems the compliance checks (e.g., ISO 26262, IEC 62304, DO178B/C) have become mandatory. Ensuring this compliance using manual analysis is error prone and costly. Therefore, many compliance checks are validated using static analysis tools. Consequently, these tools have become an integral part of software development life cycle of embedded systems. 

Most static analysis tools rely on a pointer analysis to get the actual memory locations pointed to by a pointer at its dereference points. If this underlying points-to information is accurate, it helps tools to gain more precision in their top level analysis~\cite{shapiro1997effects}. This basic demand to get precise points-to information has led to the development of several pointer analysis techniques~\cite{wilson1995efficient,andersen1994program,hardekopf2009semi,guyer2003client,sridharan2005demand,steensgaard1996points,hardekopf2007ant,pearce2007efficient,nystrom2004bottom,hardekopf2011flow,yu2010level}. Theoretically these techniques provide different levels of precision and scalability depending on the choice of flow and context sensitivity. 
In particular, a flow in-sensitive analysis does not consider the ordering of statements in a program while computing the points-to information. In contrast, the ordering of statements is considered by a flow sensitive analysis. Additionally, a context sensitive analysis considers the call sequence of functions while computing  the points-to information along with the ordering of statements within the functions.
In general, it is assumed that precision increases when we move from flow insensitive analysis to flow sensitive analysis to context sensitive analysis, while scalability decreases in the same order.


We have developed a static analysis tool~\cite{TCSECA} to detect bugs like \textit{integer overflow-underflow, division by zero}, and \textit{array index out of bound}. While analyzing few embedded industry code sets, we failed to see significant improvement in the precision of the tool when using context sensitive pointer analysis in place of flow sensitive analysis. On further investigation, we found that there was no significant difference in precision of the points-to information itself. This lead to the question ``\textit{Is this behavior peculiar to the code sets we were analyzing, or does it hold in most embedded code sets}?". To resolve this, we performed experiments and compared precision of the three pointer analysis techniques\textemdash flow sensitive, flow insensitive, and context sensitive \textemdash on seven diverse embedded industry code sets. 

 Our results show that on average in 90\% cases all three analyses (mentioned in the previous paragraph) had equal precision. Moreover, these equal precision cases varied from 76\% to 100\% on different code sets. In particular, we found that the relative gain in precision is not only dependent on the technique but is also dependent on the code. In this paper, we detail the observed differences in precision based on code pattern and correlate the two.


Given that pointer analysis has been an active area of research for several years, similar studies exist in the literature~\cite{hind2001pointer,lhotak2006context,hind2001evaluating,ruf1995context,shapiro1997effects,hind1998assessing,engblom1999static}. However, we did not find any case study that targets embedded code sets used in industry, and reasons about the difference in precision based on patterns in embedded code. Mainly because embedded industry code sets are not openly available. 
We believe such case studies play a key part in designing highly targeted analysis that are more effective than general ones. The need for such targeted analysis has been reiterated and discussed~\cite{guyer2003client,hind2001pointer,sridharan2005demand,guyer2005error} several times.

The main contributions of this paper are:
\begin{itemize}
	\item Comparison of three pointer analysis techniques on seven diverse embedded industry codes, ranging in size from 3 KLOC to 59 KLOC.
	\item Detailed analysis of correlation between embedded code patterns and pointer analysis effectiveness.
\end{itemize}

Rest of the paper is organized as follows: Section~\ref{sec:Background} details the pointer analysis techniques that we compared. Section~\ref{sec:setup} details the experimental setup. Section~\ref{sec:Results} discusses results and directions for future research. Section~\ref{sec:Related Work} presents related work and Section~\ref{sec:Conclusion} concludes the paper.

\section{Background}\label{sec:Background}
\newcommand{\langc}{\text{$\mathcal{C}$}\xspace}
\newcommand{\allp}{\text{$\mathcal{P}$}\xspace}
\newcommand{\allv}{\text{$\mathcal{V}$}\xspace}
\newcommand{\setsym}[1]{\text{$\mathcal{#1}$}\xspace}
\newcommand{\inn}[1]{\text{IN$_{#1}$}\xspace}
\newcommand{\outn}[1]{\text{OUT$_{#1}$}\xspace}
\newcommand{\genn}[1]{\text{GEN$_{#1}$}\xspace}
\newcommand{\killn}[1]{\text{KILL$_{#1}$}\xspace}

\newcommand{\ptras}{\text{\emph{PtrAssignments}}\xspace}
In this section, we explain the model used and the pointer analysis techniques that we compared.
\subsection{Model And Abstraction}
We model our approaches for the pointer analysis as applicable to \langc programs.
In \langc, any modification of the points-to information of a pointer can be realized from the following statements: address-of (p=\&a), copy (p=q), load (p=*q), and store (*p=q). Henceforth, we  call such statements as  \textit{PtrAssignments}. 

In \langc, we allocate heap memory using either \emph{malloc, calloc,} or \emph{alloc}. Here, we abstract heap by callsite i.e., we do not distinguish between different heap allocations that can happen by same allocation call (for example in presence of loops or recursion). In particular, a heap allocation call is abstractly  represented by the line number of the allocation call in the program, for example, the memory allocation assignment ``ptr = malloc" is modeled as ``ptr = \&Heap$_\mathrm{\emph{\#line}}$", where \emph{\#line} is the line number of the assignment. Moreover, an array is considered as a monolithic scalar entity, where we do not distinguish between different elements of the same array. Next, we  disassemble the aggregate types to their component fields (field-sensitive). Further, uninitialized global declarations are considered to be initialized to 0 (\textit{null}), and uninitialized locals are initialized to special value \textit{unknown}. 

We represent the application being analyzed by a 
directed graph called the Program Call Graph (PCG) with the application's entry function (typically \textit{main}) as the start node of the graph. 
Additionally each node in the graph represents a function in the application and directed edges are present between caller and corresponding called functions. Moreover, each function's body is represented by a separate control flow graph (CFG).

We use the following representation for points-to information. At each program point, we store the set of possible memory locations pointed by each pointer i.e.,
let $\allp$ be set of all pointers in the program, and 
	$\allv$ be set of all possible memory locations that could be pointed to in the program
then, 
PointsToInfo (PI): $\allp\mapsto2^\allv$ 

\subsection{Approach Description}
We now describe the details of flow insensitive, flow sensitive, and context sensitive pointer analysis below. 

\textbf{Flow Insensitive Pointer Analysis}:
Here, flow insensitive analysis refers to flow and context insensitive analysis. In this, we perform Anderson's~\cite{andersen1994program} flow insensitive pointer analysis iteratively, where  all \ptras in the program are iteratively solved and a directed Points-to graph is constructed. The graph contains two types of nodes namely pointer nodes (PtrN) and pointee nodes (PtdN). A PtrN  represents a pointer in the program, whereas PtdN represents a set of pointees (i.e., a set of memory locations). To resolve multilevel pointers, we iterate until the graph reaches a fixed point i.e., the information at each node does not change.


\textbf{Flow Sensitive Pointer Analysis}: Here, flow sensitive analysis refers to  flow sensitive and context insensitive analysis. 
In this, we use functional approach of inter-procedural analysis~\cite{sharir1978two}  for computing points-to information. The approach involves two phases. In the first phase, we iterate over
all functions in the PCG and create a summary for each of the functions. 
In the second phase, we analyze each function to compute the following two information for all nodes \textit{n} in the CFG of the function until a fix-point (saturation) is reached: let IN$_n$ and OUT$_n$ represent the program points just before and after the execution of a node $n$, and pred(n) be the set of immediate predecessors of node $n$ in the CFG of the function. IN$_{n}$ for the start node of the CFG is a special value called \textit{Boundary}. 
\begin{align}
\inn{n}&=\begin{cases}Boundary & \text{n=start node}\\
 \displaystyle\bigcup_{x\in pred(n)} \outn{x}& \text{otherwise}\end{cases}\\
\outn{n}&=\genn{n} \cup (\inn{n}-\killn{n})
\end{align}

Table \ref{transTable} summarizes the GEN and KILL information for each of the \ptras node type.  The information at the exit node of the CFG of a function is called the summary of the function.

A function boundary represents the information at the IN of start node of the function CFG. The boundary of the entry function (typically \emph{main}) is computed by processing all the global declarations in the program. For other functions the initial boundary value is a special value called Top ($\top$). When the call node $c$ of a function $f$ is encountered then the meet (typically union) of $f$'s existing boundary value and the value present in IN$_\mathrm{\textit{c}}$  is set as the new boundary of $f$. The function summary of $f$ (that has GEN and KILL information) is used to compute OUT$_\mathrm{\textit{c}}$. We keep iterating over functions in the PCG as long as there is change in either the information at any node, the function boundary, or the summary of any function.
\newcommand{\fone}{\text{$f_1$}\xspace}
\newcommand{\ftwo}{\text{$f_2$}\xspace}
\newcommand*{\MyIndent}{\hspace*{0.25cm}}
\input{table1}

\textbf{Context Sensitive Pointer Analysis}: Here, the context sensitive analysis refers to the flow and context sensitive analysis. Additionally, a \emph{calling context} refers to the data flow value at the IN of a call point of a function. Here, we solve every called function separately for each of its distinct calling context. Two calling contexts of a function are distinct if they have different data flow values~\cite{padhye2013interprocedural}. We compute the function summary for each distinct context and store $\langle$\emph{function, context, summary}$\rangle$ tuple. We start solving from the entry function with all the global initializations as its context. While solving any function \fone, when a call node \emph{c} is encountered, IN$_\mathrm{\textit{c}}$ is used to compute context \emph{ctx} of the called function \ftwo. If tuple $\langle$\emph{\ftwo, ctx, sum}$\rangle$ exists then \emph{sum} is used to compute Out$_\mathrm{\textit{c}}$, else \fone's solving is suspended till the summary is computed for \ftwo w.r.t context \emph{ctx}. We iterate over the PCG as long as either new contexts are added/updated or for any context, information changes at any node or  a function summary changes. 

\section{Experimental Setup}\label{sec:setup}
We performed our experiments on an \textit{intel core i-7, 8 GB RAM, windows 64 bit } machine. We implemented the three pointer analysis techniques in our inhouse TCS Embedded Code Analyzer(TCS ECA) Tool~\cite{TCSECA}. In this, we included seven diverse proprietary embedded code sets from the industry. These code sets were developed by different manufacturers from the automotive industry, different teams within different organizations. Few of these code sets are safety-critical systems. 
Table \ref{app_characteristics} presents the characteristics of these code sets. In particular, the size of code sets range from 3 to 59 KLOC, while the number of functions are in the range of 30 to 790. Only one of the code set (i.e., CAN1) contains heap allocation calls, and 3 code sets have 2 to 25 function pointer dereferences. Overall pointer dereference instances range from 52 to 3047 across code sets.


\input{table2}



 We treat each level of a multi-level pointer dereference as a different Point of Interest (PoI). Subsequently, for each PoI where a pointer \textit{ptr} is dereferenced, we compute points-to information \textit{ptdSet}  corresponding to the three pointer analyses:
\begin{itemize}
\item Flow Insensitive Information (FIS) = \textit{ptdSet$_{FIS}$}
\item Flow Sensitive Information (FS) = \textit{ptdSet$_{FS}$}
\item Context Sensitive Information (CS) = \{ \textit{ptdSet$_{CSi}$} \}, where \textit{i} denotes the information corresponding to the \textit{i$^{th}$} context.
\end{itemize}
A context sensitive information for \textit{ptr} is considered be more precise, if at least one context has more precise information compared to the corresponding FS or FIS information at the point (this includes the added precision when the client analysis can enquire points-to information in specific contexts). We compare FS, FIS, and CS informations for each \textit{ptr} and bucket the judgment in one of the following pointee-relation classes:
\begin{itemize}
\setlength\itemsep{7pt}
	\item \textbf{$FIS=FS=CS$}: points-to information of \textit{ptr} is same in FIS, FS, and CS i.e.,
	
				$ \forall i.\ ( ptdSet_{FIS} \equiv ptdSet_{FS} \equiv ptdSet_{CSi} ) $				
	\item \textbf{$ FIS < FS < CS $}: points-to information of \textit{ptr} in: 1) CS is more precise than FS, and 2) FS is more precise than FIS i.e.,
	
				 $ \exists i.\ (ptdSet_{CSi} \subset  ptdSet_{FS} ) \land (ptdSet_{FS} \subset  ptdSet_{FIS}) $
		\item \textbf{$ FIS = FS < CS $}: points-to information of \textit{ptr} is same in FS and FIS, and is less precise than CS i.e.,
		
		$ \exists i.\ (ptdSet_{CSi} \subset  ptdSet_{FS} )  \land (ptdSet_{FIS} \equiv  ptdSet_{FS}) $
		\item \textbf{$ FIS < FS = CS $}: points-to information of \textit{ptr} in CS and FS is same, and is more precise than FIS i.e.,
		
		$ \forall i.\ (ptdSet_{FS} \equiv  ptdSet_{CSi} ) \land (ptdSet_{FS} \subset  ptdSet_{FIS}) $
\end{itemize}

Observe that the above cases are mutually exclusive and exhaustive (i.e., cover all possible scenarios). More specifically, the result of FIS is an over-approximaton of result of FS which is an over-approximation  of CS (since CS is flow and context sensitive). Therefore, cases where FIS is more precise than FS, or FS is more precise than CS are not possible.

\section{Results and Discussion}\label{sec:Results}
Table \ref{result_table1} presents the results of our experiments, and summarizes the distribution of the PoI's into pointee-relation classes. More specifically, in 5 out of 7 applications, more than 90\% PoI had equal precision (i.e., \textit{FIS=FS=CS}), for rest two application (ATCM and CAN1) there was a precision gain of 16\% and 20\% while moving from flow insensitive to flow sensitive  analysis.

We analyzed the coding patterns, the type of pointer (constant, formal, global or local), and the type of the pointer assignments (single assignment, multiple assignment) affecting the precision. Below we describe the concrete \% of occurrence of each of these patterns that led to these results.

	
1.\textbf{$FIS=FS=CS:$} We observed that on average in 90.5\% cases the flow sensitive, the flow insensitive, and the context sensitive analysis gave the same results. This result can be explained by ubiquitous presence of the following code patterns in the code we analyzed. 

\begin{itemize}
	\item \textit{Constant Pointer}: 
	In this case, a pointer is declared as constant pointer and is assigned a memory address in declaration. For such case, it is 
	intuitive that \textit{FIS=FS=CS}. In the embedded code sets, information is read and written to hard coded memory addresses through constant pointers or via dereference of constant address directly. These cases contributed on average 33.7\% POIs in \textit{FIS=FS=CS} cases, and 31\% POIs in overall cases.
	
	\item \textit{Formal Pointer}: 
    A formal pointer refers to a pointer declared as a function parameter. 
    30.7\% cases of \textit{FIS=FS=CS} category are of the formal pointer dereferences. In this, we observed that the formal pointers are not reassigned in the procedure. Consequently, the points-to information for the flow insensitive and the flow sensitive analysis is same (i.e., FIS=FS) for such pointer dereferences within the function.
	
Further, context sensitive analysis also does not add any value in these cases because the formal pointer is assigned the same actual parameter at all the call sites of the function. This results in the same points-to information in any context as that of FS and FIS. 4 out of 7  applications depict this behavior where formals are assigned to the same variable at all call sites. Interestingly, in many procedures formal pointers are used to manipulate the values of particular global variables/arrays through pointer mimicking the setter/getter functions in Java.
	
	\item \textit{Single Assigned Pointers}: pointers are assigned only once in the entire program and the assignment occurs before the dereference point, so points-to information is same in \textit{FIS, FS,} and \textit{CS}. We observed 16.2\% cases among the \textit{FIS=FS=CS} cases are of this type. 
	
	\item \textit{Multi Assigned Pointers}: We observed that in 19.4\% cases among \textit{FIS=FS=CS} cases were under this category, where a pointer is assigned multiple values at different points and all definitions reach to the dereference point through at least one path which results in \textit{FIS=FS=CS}.  
\end{itemize}

\input{table3}

Figure \ref{equal_distribution_tbl} presents the percentage of: Constant Pointer, Formal Pointer, Single Assignment, and Multiple Assignment in the individual applications. It also highlights that none of the pattern is prevalent across the applications, yet the overall impact on the precision is same.
\input{figure1}
2.\textbf{$FIS<FS<CS:$} This category measures cases where the flow sensitivity and the context sensitivity both add some disjoint precision to the results. Even though we expected this case to be prevalent, we found only 1 out of 5403 analyzed cases of this type. 
In this lone case, a formal pointer was assigned different addresses at different callsites and it was conditionally updated inside an \textit{if-then} branch in the procedure after which it was dereferenced in the same procedure. The calls from multiple sites with different addresses made the context sensitive analysis more precise than FIS and FS, while the re-assignment of formal pointer\textemdash\textit{ which we never saw happening anywhere else in these code sets}\textemdash in the \textit{if-then} branch made FS more precise than FIS for dereference in the \textit{else} branch.
The dearth of cases in this category can be attributed to practice of not re-assigning formal pointers in the corresponding procedures.

3.\textbf{$FIS=FS<CS:$} Given that we observed only 1 case of $FIS<FS<CS$ category, we can say $FIS=FS<CS$ category measures the complete impact of the context sensitivity on the precision. Here, we observed an average of 2.6\% cases of this type. This indicates that context sensitivity has not added much value over a flow sensitive analysis across the code sets. One of the reason for this is that in embedded code sets, most formal pointers are used for manipulation of \textit{specific} variables, meaning they will be assigned same variable at all callsites. Consequently, a context sensitive analysis over such code may not improve precision. 
On the other hand, we did observe improvement in  5-6\% cases in applications namely ATCM, BTCM, and BCM where formal pointers were assigned different variables at different call sites, indicating context sensitive analysis' impact is not negligible. Nevertheless, a context sensitive analysis involves significant overhead over flow sensitive analysis like maintaining contexts, separate procedure solving for each context. In such cases, it is a good idea to do a lightweight pre-analysis to anticipate the usefulness of context sensitivity before making an analysis choice.

4.\textbf{$FIS<FS=CS:$} Given that CS (refers to context and flow sensitive) is at least as precise as FS i.e., $FS>CS$ is infeasible case, and $FIS<FS<CS$ has occurred only once, we can say $FIS<FS=CS$ measures the complete impact of the flow sensitivity on the precision. We found 6.8\% cases of this type. 
Among these cases 85\% cases are of the following type. An uninitialized global pointer is assigned at single location before use point. Since C semantics say uninitialized global pointer is initialized to \textit{null} by default, the flow insensitive points-to information for such pointers contains \textit{null} along with the actual assigned address, while the flow sensitive and the context sensitive analysis contains only the assigned address because \textit{null} is killed at the assignment location.
Apart from this, in 11\% of the cases a local pointer was initialized with \textit{null}, and it was re-assigned at single location before use, giving similar result as the above 85\% cases. Lastly, in the remaining 4\% of the cases pointer was assigned before and after its use point resulting into difference in the flow sensitive and the flow insensitive points to information at such use point.

\input{table4}
\subsection{Analysis Time}
Table~\ref{tab:analysisTime} shows the time taken by the three analyses on different code sets. In particular, the FIS analysis took 1-6 micro seconds to complete while a flow sensitive and context sensitive analysis took up to 3 and 6 seconds to complete respectively. In comparison, the context sensitive analysis took up to 200\% more time than flow sensitive analysis. On the other hand,  a flow sensitive analysis time was more than flow insensitive analysis by a magnitude of $10^6$.

\subsection{Result Summary and Future Directions}
Our code sets showcased varied amount of precision gain due to the differences in the coding constructs used. In particular, the dereference of constant pointers, and the pointers which are assigned the same pointees from different locations resulted in $FIS=FS=CS$, these cases were dominant in all the embedded code sets. Moreover, different assignments to formal pointers contributed to the $FIS=FS<CS$ bucket. Finally, most cases where $FIS<FS=CS$ was due to \langc standard where uninitialized pointers are default assigned \textit{null} . Having said that, if one has to implement a pointer analysis technique then it will be advisable to keep these trends in mind and understand the code for which the analysis has to be applied, and then make a choice of technique as per one's requirement.

More specifically, one can write a pre-analysis that scans the code to quantify the presence of the above patterns in the code. Subsequently, the output of pre-analysis can be used to automatically decide which of the three analyses to do on the code. For instance, if the code contains no cases where a formal pointer is assigned more than one variables then doing context sensitive analysis is not useful in such case.

\section{Related Work}\label{sec:Related Work}
Das et al.~\cite{das2001estimating} showcased that the flow sensitive analysis provided almost the same level of precision on 95\% cases, they showed this for large open source applications of uptill 2MLoC. Their conclusion on precision gain has been in coherence with Ruf~\cite{ruf1995context}, Foster~\cite{foster2000polymorphic}.
Ruf et al.~\cite{ruf1995context} presented an empirical study of the flow sensitive and the context sensitive analysis. They showed with caution that there can be no improvement in the precision at the dereference locations. We have shown a similar result on the embedded code sets. In our work, we also presented a reasoning based on the code patterns in the embedded code sets which was not done in the earlier work.

Engblom et al.~\cite{engblom1999static} studied embedded system's static property, they aimed to improve and guide development of \emph{worst case execution time} analysis. Whereas, our focus is on pointer usage and its assignment pattern that affects the precision of pointer analysis. They also concluded that pointer analysis is necessary to obtain good results on embedded code.
Liang et al.~\cite{liang2005evaluating} and Lhotak et al.~\cite{lhotak2006context} evaluated the effectiveness of the context sensitive points-to analysis on the Java programs. Their results show that the context sensitivity increases the precision for some benchmarks, and the context sensitive heap abstraction is important for precision.
Empirical studies by Hind et al.~\cite{hind2001evaluating,hind1998assessing} focused on comparing the flow insensitive and the flow sensitive analysis with different tuning,  both studies concluded that for most applications the two type of analysis didn't show the expected  precision gain difference. They gave intuitive reasoning for such behavior, whereas we had extended it to the code level reasoning making it more concrete.

\section{Conclusion}\label{sec:Conclusion}
This work described an empirical study carried out to evaluate the effectiveness of three different pointer analysis techniques on seven diverse embedded code sets used in industry. We found in 90\% cases the flow insensitive analysis is as precise as flow and context sensitive analysis. Subsequently, we highlighted the correlation between the precision gain and the code constructs used in the industry. 
We also discussed how the results of this study can be used to design targeted and effective pointer analyses for embedded industry code sets.

\bibliographystyle{ieeetr}
\bibliography{main}

\end{document}

%% file: table1.tex
\begin{table}[]
\centering
\caption{GEN and KILL for the node types in \ptras}
\label{transTable}
\begin{tabular}{|c| l |}
\hline
Node \textit{n} & \begin{tabular}{@{}c@{}} Output Action At a Node  \end{tabular} \\ \hline \hline
 \begin{tabular}{@{}c@{}} Address-of \\ p=\&a \\ \end{tabular} & 
 \begin{tabular}{@{}l@{}} KILL$_\mathrm{\textit{n}}$ = \{p\} \\
					      \text{// p points-to a }\\
								GEN$_\mathrm{\textit{n}}$ = \{ (p, \{a\}) \}
								\end{tabular} 
 \\ 
 \hline
 \begin{tabular}{@{}c@{}} Copy \\ p=q  \\ \end{tabular} &
 \begin{tabular}{@{}l@{}} KILL$_\mathrm{\textit{n}}$ = \{p\} \\
								\text{// p points-to pointees of q}\\
								GEN$_\mathrm{\textit{n}}$ = \{ $ (p, \setsym{Y}) \mid (q, \setsym{Y}) \in$  IN$_\mathrm{\textit{n}}$ \}
								
 \end{tabular}
 \\
 \hline
 \begin{tabular}{@{}c@{}} Load \\ p=*q  \\ \end{tabular} &  
  \begin{tabular}{@{}l@{}}  KILL$_\mathrm{\textit{n}}$ = \{p\} \\
	\text{// p points-to pointees of pointees of q}\\
						  GEN$_\mathrm{\textit{n}}$ = \{ (p, \setsym{S}) \}, where\\
							\setsym{S} = $\bigcup_{x \in \text{pointee}_q}$ \{ y $\mid$  y $\in$ \setsym{S'} $ \wedge $ ( x, \setsym{S'}) $\in$ IN$_\mathrm{\textit{n}}$  \} \\
							pointee$_q$ = \{ x $\mid$  x $\in$ \setsym{S''} $\wedge\ (q, \setsym{S''})$ $\in$  IN$_\mathrm{\textit{n}}$   \} 
							
						 \end{tabular} \\ \hline
 \begin{tabular}{@{}c@{}}Store \\ *p=q \\ \end{tabular} & 
\begin{tabular}{@{}l@{}} pointee$_p$ = \{ x $\mid$ x $\in$ \setsym{S} $ \wedge $ (p, \setsym{S}) $\in$  IN$_\mathrm{\textit{n}}$ \} \\
\text{// if single pointee then kill }\\
					 	KILL$_\mathrm{\textit{n}}$ = ( $|$pointee$_p$$|$ == 1) ? pointee$_p$ : \{ \} \\
					  \text{// pointees of p points-to pointees of q}\\
						GEN$_\mathrm{\textit{n}}$ = \{ (x, \setsym{Y}) $\mid$ x $\in$ pointee$_p$ $ \wedge $ (q, \setsym{Y}) $\in$  IN$_\mathrm{\textit{n}}$   \}
							
					 \end{tabular} \\ \hline	
\end{tabular}
\end{table}

%% file: table2.tex
\begin{table*}[]
\centering
\caption{Application Characteristics-1}
\label{app_characteristics}
\begin{tabular}{|l|r|r|r|r|r|r|r|r|r|r|r|r|r|}
\hline
\multicolumn{1}{|c|}{\multirow{3}{*}{}} & 
\multicolumn{1}{c|}{} & 
\multicolumn{2}{c|}{\multirow{2}{*}{\#Nodes}} & \multicolumn{4}{c|}{\#Pointer Assignment Nodes} 
&&&&&&
\\ \cline{5-8}

\multicolumn{1}{|c|}{} & 
\multicolumn{1}{c|}{} & 
\multicolumn{1}{c}{} & 
\multicolumn{1}{c|}{} & \multicolumn{2}{c|}{Initialization} & \multicolumn{2}{c|}{Assignment}  

&\multicolumn{1}{c|}{\begin{tabular}[c]{@{}c@{}}\#Heap\\ Allocation\\ Sites\end{tabular}} & \multicolumn{1}{c|}{\begin{tabular}[c]{@{}c@{}}\#Function\\ Pointer\\ Derefs\end{tabular}} & \multicolumn{1}{c|}{\begin{tabular}[c]{@{}c@{}}\#Funcs\\ Called\\ Once\end{tabular}} & \multicolumn{1}{c|}{\begin{tabular}[c]{@{}c@{}}\#Funcs\\ Called\\ More\\ Than \\ Once\end{tabular}} & \multicolumn{1}{c|}{\begin{tabular}[c]{@{}c@{}}\#Total\\ Derefs\end{tabular}} & \multicolumn{1}{c|}{\begin{tabular}[c]{@{}c@{}}\#Derefs \\ Resolving\\ to\\ Single \\ Location\end{tabular}}

\\ \cline{3-8}

\multicolumn{1}{|c|}{\rotatebox{90}{Application}} & \multicolumn{1}{c|}{\rotatebox{90}{KLoC}} & 
\multicolumn{1}{c|}{PCG} & 
\multicolumn{1}{c|}{CFG} & 
\multicolumn{1}{c|}{\rotatebox{90}{Local}} & \multicolumn{1}{c|}{\rotatebox{90}{Global}} & \multicolumn{1}{c|}{\rotatebox{90}{Local}} & \multicolumn{1}{c|}{\begin{tabular}[c]{@{}c@{}}Actual\\ to\\ Formal\end{tabular}} 
&&&&&&
\\ \hline

SMCARD &  3 & 48 & 1929 & 2 & 37 & 0 & 254 
& 0 & 0 & 23 & 25 & 52 & 52\\ \hline

NavPS &  33 & 66 & 1956& 1 & 86  & 0 & 21 
& 0 & 0 & 47 & 19 & 426 & 419\\ \hline

ATCM & 18 & 235 & 4146 & 0 & 31  & 89 & 43 
& 0 & 11 & 143 & 92 & 552 & 351\\ \hline

CAN1 & 30 & 30 & 1749 & 37 & 135  & 140 & 286 
& 26 & 2 & 23 & 7 & 627 & 462\\ \hline

BTCM &  40 & 923 & 27588 & 28 & 242  & 84 & 395 
& 0 & 0 & 855 & 68 & 3047 & 2870\\ \hline

BCM &  59 & 790 & 23565 & 2 & 377  & 75 & 184 
& 0 & 25 & 607 & 183 & 178 & 121\\ \hline

CAN2 & 37 & 125 & 2640 & 6 & 36  & 26 & 39 
& 0 & 0 & 109 & 16 & 520 & 351\\ \hline
\end{tabular}
\end{table*} 

%% file: table3.tex
\begin{table}[t]
\centering
\caption{Pointer Analysis Results: Abbreviations used- FIS$\rightarrow$ \textit{flow insensitive}, FS$\rightarrow$ \textit{flow sensitive}, CS$\rightarrow$ \textit{context sensitive}}
\label{result_table1}
\begin{tabular}{|l|r|r|r|r|}
\cline{2-5}
  \multicolumn{1}{c|}{\multirow{2}{*}{}}
  & \multicolumn{1}{c|}{\multirow{2}{*}{\begin{tabular}[c]{@{}c@{}}Total\\ PoIs\end{tabular}}} &
  \multicolumn{3}{c|}{\#PoIs(\%)}
  \\ \cline{3-5}
  
  \multicolumn{1}{c|}{\multirow{2}{*}{}}
  & \multicolumn{1}{c|}{} &
  \multicolumn{1}{l|}{\begin{tabular}[c]{@{}c@{}}
  FIS = FS,\\ FS = CS\end{tabular}} &
  \multicolumn{1}{l|}{\begin{tabular}[c]{@{}c@{}}
  FIS = FS,\\ FS $<$ CS\end{tabular}} & \multicolumn{1}{l|}{\begin{tabular}[c]{@{}c@{}}
  FIS $<$ FS,\\ FS = CS\end{tabular}} \\ \hline
 
  SMCARD & 52 & 52 (100) & 0 (0) & 0 (0) \\ \hline 
  NavPS & 426 & 424 (99.5) & 2 (0.5) & 0 (0) \\ \hline
  ATCM & 552 & 400 (72.4) & 28 (5) & 124 (22.6) \\ \hline 
  CAN1 & 627 & 515 (82.1) & 12 (1.9) & 100 (16) \\ \hline 
  BTCM & 3047 & 2895 (95) & 150 (4.93) & 2 (0.07) \\ \hline 
  BCM & 178 & 161 (90.4) & 11 (6.2) & 6 (3.4) \\ \hline 
  CAN2 & 520 & 489 (94.1) & 0 (0) & 31 (5.9) \\ \hline\hline
\multicolumn{1}{|l|}{Average}  & 771.7 & 705 ( 90.52) & 29 (2.65) & 37.57 (6.83) \\ \hline
\end{tabular}
\end{table}

%% file: figure1.tex
\definecolor{my_blue}{RGB}{135, 206, 250}
\begin{figure*}[t]
\centering
\caption{Application Wise Distribution of Code Pattern for \textit{FIS = FS = CS}}
\label{equal_distribution_tbl}
\begin{tikzpicture}[thick, scale=0.67]
\begin{axis}[
    axis lines=left,
    width=\linewidth *1.5,
    height=4.0cm,
    ybar,
		area legend,
    enlargelimits=0.10,
    enlarge y limits={300,value=0.1},
    legend style={at={(.4, -0.4)}, anchor=north,legend columns=-1},
    ylabel={\%},
    xlabel={Application},
    symbolic x coords={SMCARD, NavPS, ATCM, CAN1, BTCM, BCM, CAN2},
    xtick=data,
    nodes near coords,
    nodes near coords align={vertical},
    ]
\addplot [fill=red, postaction={pattern=horizontal lines}, bar shift=-0.8cm, ] 
	coordinates {(SMCARD,0) (NavPS,97.2) (ATCM,47.5) (CAN1,0) (BTCM,57.8) (BCM,0) (CAN2,34.4)};
\addplot [fill=my_blue, postaction={pattern=north east lines}, bar shift=-0.25cm, ]
	coordinates {(SMCARD,100) (NavPS,2.8) (ATCM,35.25) (CAN1,19) (BTCM,7.4) (BCM,15.5) (CAN2,35)};
	\addplot [fill=yellow, postaction={pattern=vertical lines}, bar shift=0.25cm, ] 
	coordinates {(SMCARD,0) (NavPS,0) (ATCM,17) (CAN1,5.4) (BTCM,10.8) (BCM,75.1) (CAN2,3.7)};
\addplot [fill=green, postaction={pattern=north west lines}, bar shift=0.8cm, ]
	coordinates {(SMCARD,0) (NavPS,0) (ATCM,0.25) (CAN1,75.6) (BTCM,24) (BCM,9.4) (CAN2,26.8)};

\legend{Const Pointer, Formal Pointer, Single Assignment, Multiple Assignment}
\end{axis}
\end{tikzpicture}
\end{figure*}

%% file: table4.tex
\begin{table}[t]
\centering
\caption{Analysis time in seconds. The FIS analysis took up to 6 microseconds, while FS and CS analysis took up to 3 and 6 seconds respectively.}
\label{tab:analysisTime}
\begin{tabular}{|l|r|r|r|}
\cline{2-4}
  
  \multicolumn{1}{c|}{\multirow{2}{*}{}}
  & \multicolumn{1}{l|}{
  FIS} &
  \multicolumn{1}{l|}{FS} & 
  \multicolumn{1}{l|}{CS} \\ \hline
 
  SMCARD & $1*10^{-6}$ & 0.4  & 0.6  \\ \hline 
  NavPS & $5.7*10^{-6}$  & 1.14 & 3.18  \\ \hline
  ATCM & $6.7*10^{-6}$  & 1.06 & 1.43  \\ \hline 
  CAN1 & $4.2*10^{-6}$  & 1.14 & 1.81  \\ \hline 
  BTCM & $13*10^{-6}$  & 3 & 6 \\ \hline 
  BCM & $3*10^{-6}$  & 0.6 & 1.1   \\ \hline 
  CAN2 & $4.5*10^{-6}$  & 0.87 & 1.07  \\ \hline
\end{tabular}
\end{table}